\documentstyle[aps,prd,epsfig,floats]{revtex} 
\draft

\newcounter{savefig}

\begin{document}
\twocolumn[\hsize\textwidth\columnwidth\hsize\csname
@twocolumnfalse\endcsname
\title{%
\hbox to\hsize{\normalsize\rm May 2002
\hfil Preprint MPI-PTh/2002-21}
\vskip 36pt Design of Adiabatic Logic for a Quantum CNOT Gate}

\author{V. Corato}
\address{Istituto di Cibernetica del CNR, via Campi Flegrei 34,
I-80078, Pozzuoli, Napoli, Italy}
\author{P. Silvestrini}
\address{Second University of Naples, Via Roma 29, I-81031 Aversa,
Italy and
Istituto di Cibernetica del CNR, via Campi Flegrei 34, I-80078,
Pozzuoli, Italy}

\author{L.~Stodolsky}
\address{Max-Planck-Institut f\"ur Physik 
(Werner-Heisenberg-Institut),
F\"ohringer Ring 6, 80805 M\"unchen, Germany}

\author{J. Wosiek}
\address{M. Smoluchowksi Institute of Physics, Jagellonian
University, Reymonta 4, 30-059 Cracow, Poland}

\maketitle

\begin{abstract} 
We examine the realization of a quantum CNOT gate by adiabatic
operations.   
The principles of such systems and their analysis are briefly
discussed and 
a model  consisting of  two weakly coupled double-
potential well qubits is studied numerically. Regions
of the parameter space with suitable  well-defined sets of
wavefunctions are found, in which then an adiabatic sweep of
an external bias produces the switching behavior of  CNOT.
Results are  presented on the adiabatic condition  and the
identification with
the parameters of a flux-coupled two-SQUID system is given.
For typical parameters  adiabatic times in the
nanosecond regime are obtained.
 
\end{abstract}
\vskip2.0pc]

The basic element of the ``quantum computer'' \cite{qcomp}
 is the quantum bit (qubit), a  two level system, exhibiting
quantum
coherence between the states. Many physical realizations of the
qubit have been proposed \cite{mqc2}. To manipulate the qubit
quantum gates \cite{bar95}
are necessary, logic devices capable of operating on linear
combinations of input states.
First there is the simple NOT, a one bit operation which can be
viewed as an inversion operation on a qubit. The next step is 
to construct gates of a conditional character. A simple case
to consider is the  two-bit operation 
``controlled NOT'' or CNOT. To realize such device  it is natural
to consider using an interaction
between the physical elements constituting the qubit. 

Among the possible mechanisms for manipulating coupled qubits
adiabatic procedures are, as explained below, of special interest.
Furthermore, it has been suggested  that adiabatic procedures
may be robust with respect to certain kinds of errors \cite{error}.
In particular with superconducting devices, Averin~\cite{averin} 
 has suggested using
small Josephson junctions in the coulomb blockade regime, and we
have mentioned the possibility of using SQUID qubits with flux
coupling~\cite{deco}.
In this letter we will explain some general principles for studying
such systems and to present numerical calculations relevant to
their behavior and design.

CNOT is  a two-qubit operation and we will represent it 
by two interacting  double-potential well systems. Each double well
system may be though of as an approximately independent qubit since
we  shall keep the coupling  weak. Qualitatively, we will use the
procedure of performing an adiabatic  NOT~\cite{deco} on the first
qubit while trying to influence its behavior by the state of the
second.
We   find a region
of parameter space where this works.

{\it Hamiltonian:} We  take the following model
hamiltonian
\begin{equation}
\label{ham1}
H={-1\over 2\mu_1}{\partial ^2\over\partial \phi^2_1 }+{-1\over2
\mu_2}{\partial ^2\over\partial \phi^2_2 }+V
\end{equation}
containing dimensionless ``masses'' $\mu$ and 
dimensionless ``position'' coordinates $\phi_1,\phi_2$, while the 
potential term is 
\begin{eqnarray}\label{pot} 
V=
V_0\bigl\{{1\over 2}[l_1(\phi_1-\phi^{ext}_1)^2
+l_2(\phi_2-\phi^{ext}_2)^2~~~~~~~~~~~~~~~~\cr
-2l_{12}(\phi_2-\phi^{ext}_2)(\phi_1-
\phi^{ext}_1)]+\beta_1 f(\phi_1)+\beta_2f(\phi_2) \bigr\} \; .
\end{eqnarray}

The
function $f(\phi)$ is chosen so that a double-well potential
results
for each  $\phi$ variable. We shall use $f(\phi)=1-{1\over 2}\phi^2
+{1\over 24}\phi^4$.
$V_0$, $l_1$, $l_2$, $l_{12}$, $\beta_1$ and $\beta_2$ are
constants depending on system parameters.
The two $\phi^{ext}$ are external biases which can be adjusted or
varied to perform the  operation and to find favorable operating
points for the device.  Their values  determine the degree of
asymmetry of each double well system; when they are zero the wells
are symmetric (for $l_{12}=0$). Note that when the coupling
parameter $l_{12}$ is zero the
hamiltonian simply represents  two non-interacting systems.  Our
model
system thus consists of two weakly interacting double-potential
wells
 with externally adjustable biases $\phi^{ext}_1,\phi^{ext}_2$.  
Fig 1 shows the equipotential contours of $V(\phi_1,\phi_2)$, with
its four potential wells. 
\begin{figure}
\epsfig{bbllx=100bp,bblly=140bp,bburx=400bp,bbury=430bp,    
 file=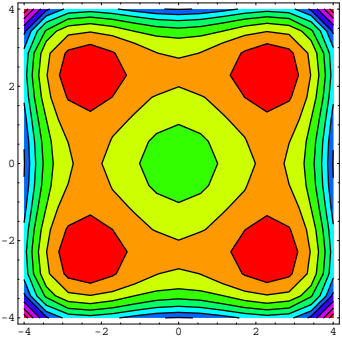, width=.55\hsize,clip=}\hfill%
\begin{minipage}[b]{.4\linewidth}
\caption{Potential as in Eq~[\protect\ref{pot}], with its four
wells. The coordinate $\phi_1$(target bit) runs horizontally and
$\phi_2$(control bit) vertically. Red indicates the deepest
potential, purple the highest.}
\end{minipage}
\end{figure}

{\it Representation of Logical States:}
We first require a representation of the four states of
the two-qubit system.  Each logical state will be
represented by a 
  wavefunction   localized in a distinct 
potential well. When this obtains, the logical state corresponds to
a distinguishable physical state with high probability. With SQUID
qubits, for
example, if for a single SQUID a wavefunction concentrated on the
left
 of the double potential well corresponds to current going
clockwise, then  for a two-SQUID system the
lower left well of Fig 1  corresponds to current counter-
clockwise in SQUID 2 and current clockwise in SQUID 1. Since, as
will be explained in the next
paragraph, we
 work with energy eigenstates,  a first requirement on the
hamiltonian  is thus that it yield a set of ``good wavefunctions'',
that is
where the first four energy eigenstates are well
localized   in the four  different wells of Fig 1.

  Assuming such  a set of  wavefunctions has been found,
visualization of the situation is aided by the use of tableaux
 indicating where the wavefunctions are localized on Fig 1.
Labeling the first
four energy eigenstates in order of increasing energy 1,2,3,4, 
examples of   possible tableaux are seen in Eq[\ref{map}].

{\it Representation of CNOT: }
 Our aim is to represent a logical operation such as CNOT by a
mapping of the set of initial logical states to a certain set of
final logical states. This will be represented by a particular
rearrangement
of states on the tableau.
 
CNOT is defined by the conditions: A)  
the control bit does not change its state, and B) the
target bit is reversed or not reversed, according to whether the
control bit is 1  or 0.
If we identify the top row of the tableau with control bit =0
 and the bottom row with
control
bit = 1,
a physical embodiment of CNOT would be 
\begin{equation} \label{map}
\pmatrix{4&3\cr1&2} \rightarrow \pmatrix{4&3\cr2&1}\; , 
\end{equation} 
 Condition A) on the stability of the control bit is exhibited in 
that no states move between the top and bottom row.
Condition B) is realized in that the top row remains unchanged
while the bottom row is ``flipped''.

{\it Adiabatic Operations: }
 Realization of operations such as Eq[\ref{map}] can be
accomplished in an especially transparent way by using adiabatic
processes. This is due to the ``no level
crossing'' behavior of adiabatic evolution.
The no-crossing property  assures that
a state initially in the first, or second, or third,.... energy
level will end up in the first, or second, or third,... energy
level after the adiabatic evolution, while at the same time the
physical properties associated
with the level may  be  changing. Thus in 
Eq[\ref{map}] with SQUIDs, state 1 begins as a configuration with
the current
clockwise in SQUID 1, counter-clockwise in SQUID 2 and ends up as
a configuration where the current  remains counter-clockwise in
SQUID 2
but is now  reversed to counter-clockwise in SQUID 1.
 
One  can proceed as follows: we search for an initial
hamiltonian whose variable parameters $(\phi_1^{ext},\phi_2^{ext})$
are adjusted to give the left tableau of Eq[\ref{map}]. Then, we
search for a final hamiltonian where another set of
$(\phi_1^{ext},\phi_2^{ext})$,   gives  the tableau on the right.
If the two parameter sets can be connected by a smooth, slow
transformation,-a ``sweep''- we have obtained an adiabatic
realization of our
operation, here CNOT.
  
In this procedure we need only to study  the  {\it stationary}
Schroedinger equation at first. This is
 an important simplification for  the numerical analysis. 
  However,  after having determined some suitable parameter sets 
we shall also study the  full time-dependent
Schroedinger equation. This is necessary
to determine what sweep speed 
 is  ``slow'', that is guarantees  adiabatic behavior.

{\it Numerical Methods:}
 Our problem involves two variables and tunneling through  four
barriers, as well as a multi-dimensional parameter space.
To deal with this complex situation
 we turn to a recently developed
method~\cite{jacek} for numerical
solution of the Schroedinger equation. A large basis of
harmonic oscillator wavefunctions is used to reduce the problem to
an array of fast algebraic manipulations,  programmed
in Mathematica. Except for the small $l_{12}\sim 10^{-3}$, we work
with
parameters of order one, hence the resulting dimensionless
energies are also of order one. However the splittings among the
lowest  levels, which are what we  manipulate, result from
tunneling and are small $\sim 10^{-3}-\sim 10^{-4}$. Hence four
place accuracy is necessary. Using these methods we have been able
to find  a 
 region of the
$(\phi_1^{ext},\phi_2^{ext})$ parameters space where there are
``good wavefunctions''. These are indicated as the gray regions of
Fig 2, with the parameters as indicated. Reducing the value of
$l_{12}$ leads to a shrinking of these regions on the plot.

As a by-product of our numerical work we can also examine
the validity of the frequently used ``psuedo-spin'' picture. One
often
usefully visualizes~\cite{deco}  the lowest quasi-degenerate levels
of the system as ``spins''. This picture requires, however, that 
the moving
statevectors
  remain in the hilbert space spanned by an initial (here four )
set of
states.  By evaluating wavefunction overlaps we find this is true,
to a good approximation,
supporting the use of the
``spin'' picture.  We stress, however,
that we do not need this simplification in our calculations.

{\it Switching Behavior:}  We have been  able to obtain 
 switching behavior according to  Eq[\ref{map}] for  the ``good''
regions of 
Fig 2, by means of the following
operation: the control bias $\phi_2^{ext}$ is held 
constant at a relatively high value while there is a  sweep of the
target bias
$\phi_1^{ext}$ . This is a generalization of a simple
NOT~\cite{deco} on $\phi_1$. The results may be understood in terms
of a simple model: the $l_{12}$ coupling produces an extra bias on
the target bit which "helps or hinders" the NOT operation. 

The relatively large bias on $\phi_2$ comes from 
 condition A): we attempt to ``immobilize'' the control bit despite
the  perturbations communicated 
by the sweep of $\phi_1^{ext}$ via $l_{12}$. We therefore
investigate the region $\vert \phi_1^{ext}\vert <<\vert
\phi_2^{ext}\vert $.
If $\phi_2$ is indeed successfully ``immobilized'' it will be
fixed in one of its two potential wells and can have only the 
values $\phi_2\approx \pm 1$.
As seen by $\phi_1$, this
amounts to $\pm$ an extra bias.
 To linear order (since we take all $\phi_1^{ext},\phi_2^{ext}$  
small
compared to $1$ and $\phi_1,\phi_2$ are in the neighborhood of $1$)
and
 introducing the notation

\begin{equation} \label{eff}
\phi_{1~eff}^{ext}=\phi_1^{ext}\pm
{l_{12}\over
l_1}
\end{equation}

 the potential terms 
involving    $\phi_1$ in Eq~[\ref{pot}] become
\begin{eqnarray} \label{effa} 
-2l_1\phi_1\phi_1^{ext}-2l_{12}\phi_1(\pm
1)~~~~~~~~~~~~~~~~~~~~~~~~~~~~~\cr =-
2l_1\phi_1(\phi_1^{ext}\pm {l_{12}\over
l_1})=-2l_1\phi_1\phi_{1~eff}^{ext}\; ,
\end{eqnarray}

As a consequence, there 
is an
effective shift in the  external bias  on $\phi_1$   by $(\pm
{l_{12}\over l_1})$. This is just as suggested by the  ``help
or hindering'' picture, and  we further learn that the magnitude of
the ``help'' is  ${l_{12}\over l_1}$. According to this
picture
we should  try to analyze the behavior of $\phi_1$
(target bit) as if it were simply under a modified  external bias 
$\phi_{1~eff}^{ext}$.

We would then expect  if  a certain
tableau obtains and  $\phi_{1}^{ext}$ is varied,  that the tableau
is maintained until there is a sign switch for 
$\phi_{1~eff}^{ext}$. To see how the sign switch occurs, note that
for $\phi_1^{ext}\approx 0$ the sign of $\phi_{1~eff}^{ext}$ is
given by
the $\pm$ from the control bit. But  for
$|\phi_1^{ext}|>>{l_{12}\over
l_1}$, the sign is controlled by $\phi_1^{ext}$  itself.
There is, therefore, a region around $\phi_1^{ext}=0$ where the
tableau is determined by the control bit, and another region for
large $\vert \phi_1^{ext} \vert$ where the tableau is determined
 by
the sign of  $\phi_1^{ext}$. As we cross from one region to
another,
one pair of states will retain the sign it had for large 
$\vert \phi_1^{ext} \vert$ and the other pair of states  will
switch. This would be the desired behavior.
According to Eq~[\ref{eff}] the point where
the switch from one tableau to another should take place is given
by $\vert \phi_1^{ext}\vert \approx l_{12}/l_1$.

{\it Numerical Results:}
 The
 three different gray areas of Fig 2 have well defined
tableaux   as follows:
\begin{equation} 
\pmatrix{3&4\cr1&2}~~~~~
\pmatrix{4&3\cr1&2}~~~~~
\pmatrix{4&3\cr2&1}\;,
\end{equation}
for the intermediate gray region (left), the darkest region
(center) and 
 the light gray region (right), respectively.
 These tableaux fit with the description arrived at in the model,
where either the top or bottom row inverts as we go from
the central region  to large $\vert \phi_1^{ext}\vert$.
Hence a sweep from the central region to the right region produces
the mapping Eq~[\ref{map}].
Similarly sweeping from the right region to
the center and from the left region  to the central region and
vice-versa  can also serve as realizations,
differing simply in the assignment of (0,1) for the quantum states
or the names for the qubits.
Fig 2 shows that the switch between
tableaux occurs quite close to 
 $\vert \phi_1^{ext}\vert =\vert l_{12}/l_1\vert$,
 as predicted by the model. Also, it seems the inequality $
\vert \phi_1^{ext}\vert <<\vert \phi_2^{ext}\vert $
need not be very strong for ``immobilization''.

{\it Adiabatic Condition:}
 An important time scale is 
$\tau_{adiab}$,  the
shortest time  in
 which an operation can be performed  adiabatically. This time is
relevant both to the maximum speed of the device and  with respect
to decoherence and relaxation effects since the operation must take
place in  times short compared to those  for these effects. We thus
now examine the {\it time dependent} Schroedinger equation
$i{\partial\over \partial \tau}\psi=H\psi$. We work with the
dimensionless time variable $\tau$, where the connection to usual
time $t$ is given  by  $\tau=E_0 t$. $E_0$ is an energy
parameter in electron volts or Hz, characteristic of
the particular system under consideration. It  also gives the
overall energy scale. Thus all energies are in units of $E_0$ and
all times are in units of $E_0^{-1}$   ($\hbar =1 $).

 An  estimate
 for NOT~\cite{deco} gave
\begin{equation} \label{tab}
\tau _{adiab} =\epsilon  \omega_{tunnel}^{-2}= \epsilon \tau
_{rabi}^2\; ,
\end{equation}
where $\epsilon$ is  the length of the sweep in energy
 and $\omega_{tunnel}^{-1}=\tau _{rabi}$,
 the inverse tunneling energy or oscillation time between the two
states at minimum separation. Since here we also perform a kind of
NOT, we expect a
smiliar relation
to hold, where  $\omega_{tunnel}$ or $\tau
_{rabi}^{-1}$ is the smallest level splitting  during the adiabatic
passage and the sweep length $\epsilon$ may be read off 
 from the total
energy shift of the wells. We  define a ``degree of
adiabaticity'' by performing a sweep numerically and taking the
overlap of the resulting wavefunction
 with the wavefunction of the  corresponding
stationary final hamiltonian; that is to say the overlap with the
wavefunction that would result from an infinitely slow sweep. The
square of this amplitude,  $P_{fi}$,   is shown in
Fig 3 as a function of sweep time, for  a
sweep   $(0.002, -0.01)\to (0.008, -0.01) $ on Fig 2.  The arrow
indicates the 
 theoretical prediction using  Eq[\ref{tab}].
As would be expected, $\tau _{adiab}\approx e^{6.2}\approx 500$ is
a large number in dimensionless units.

{\it Identification with SQUID parameters:} 
Eqns [\ref{ham1},\ref{pot}] arise   by standard methods in the
analysis of two rf SQUID
loops coupled by a mutual inductance $L_{12}$~\cite{big}. The
Josephson relation leads to  $f(\phi)=cos\phi$, to
which our $f(\phi)=1-{1\over 2}\phi^2
+{1\over 24}\phi^4$ is a good approximation.

With $L_1$, $L_2$ the SQUID inductances and $C_1$, $C_2$ the
Josephson junction capacitances, the energy
scale factor is $E_0=1/\sqrt{ LC}$, where $C=\sqrt{C_1C_2}$ and
$
L={L_1L_2-L^2_{12} \over \sqrt{L_1L_2}
}\approx \sqrt{L_1L_2} $.
A set of
reasonable values for the SQUID is
$L_1=300pH$, $L_2=280pH$, $L_{12}=1.8pH$, $C_1=C_2=0.1pF$ and
$\beta_{1}=\beta_{2}=1.28$. Since in frequency units one finds
$E_0\approx
{1\over\sqrt{L/pH ~C/pF}}~ 1000~ GHZ$ these values give $E_0\approx
185~ GHZ$. Then the  $\tau_{adiab}\approx 500$ in dimensionless
units corresponds to   $\tau_{adiab}\approx 500 /E_0\approx
 2.7\cdot
~10^{-9} s$ in seconds.

Finally we note that here, as with all discussions of  quantum
computation, the important open question is the time scale  for
decoherence 
$\tau_{dec}$. In ~\cite{deco} we  estimated  $\tau_{dec}\approx
(few)\times 10 ^{-6} s$ for the SQUID  at 40mK
 (and also suggested a
method for its direct measurement). With the above estimate for
$\tau_{adiab}$ it thus
appears, at least for the SQUID, that the
 adiabatic condition allows for  operations in times less
than
$\tau_{dec}$. 
 However, the question of
the decoherence time is controversial and system-dependent and will
probably only be
resolved convincingly by experiment. 
In this respect, it is encouraging that in experiments~\cite{squid}
showing evidence for macroscopic quantum behaviour of the
SQUID the dissipation value is rather small.

\begin{figure}
  \epsfig{file=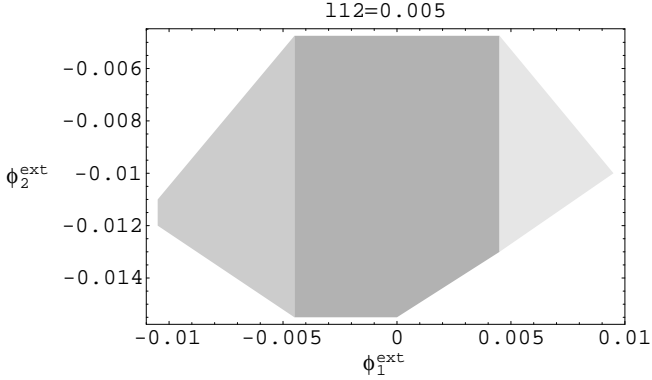, width=\hsize}
\caption{  A region of the $\phi_1^{ext},\phi_2^{ext}$ plane with
a
well defined set of wavefunctions as explained in the text. The
coupling  parameter is $l_{12}=0.005$. Other parameters  were
$l_1=l_2=1,~
\beta_1=\beta_2=1.19,~ \mu_1=\mu_2=V_0= 16.3 $}
\end{figure}

\begin{figure}
\epsfig{file=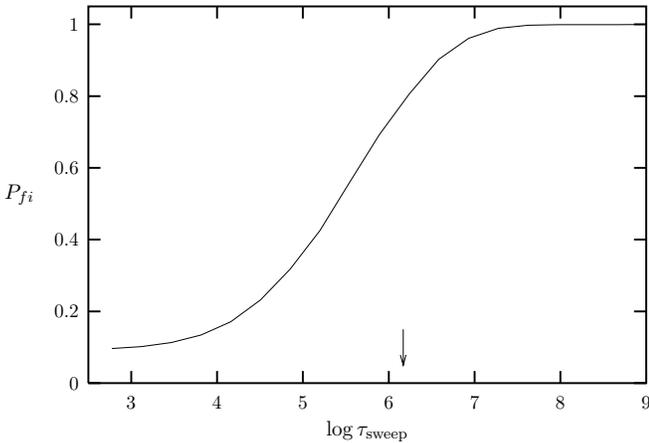, width=\hsize}
\caption{The adiabaticity parameter versus sweep time for a sweep
of ($\phi_1^{ext},\phi_2^{ext}$): 
$(0.002, -0.01)\to (0.008, -0.01) $. $P_{fi}$=1 denotes
perfect adiabaticity. The arrow indicates the theoretical estimate 
 $\tau _{adiab} =\epsilon  \omega_{tunnel}^{-2}$.}
\end{figure}

\end{document}